\documentclass[]{article}
\usepackage{cite}
\usepackage{amsmath,amssymb,amsfonts}
\usepackage{algorithmic}
\usepackage{graphicx}
\usepackage{textcomp}
\usepackage{xcolor}

\begin{document}

\title{Partitioned Deep Learning of Fluid-Structure Interaction}
\date{}

\author{Amin Totounferoush, Axel Schumacher \& Miriam Schulte \\
Institute for Parallel and Distributed Systems, Department of Computer Science\\
University of Stuttgart\\
70569 Stuttgart, Germany \\
\{amin.totounferoush, miriam.schulte\}@ipvs.uni-stuttgart.de \\ axel.schumacher@meisterfliesen.ws \\
}

\maketitle

\begin{abstract}
We present a partitioned neural network-based framework for learning of fluid-structure interaction (FSI) problems. We decompose the simulation domain into two smaller sub-domains, i.e., fluid and solid domains, and incorporate an independent neural network for each. A library is used to couple the two networks which takes care of boundary data communication, data mapping and equation coupling. Simulation data are used for training of the both neural networks. We use a combination of convolutional and recurrent neural networks (CNN and RNN) to account for both spatial and temporal connectivity. A quasi-Newton method is used to accelerate the FSI coupling convergence. We observe a very good agreement between the results of the presented framework and the classical numerical methods for simulation of 1d fluid flow inside an elastic tube. This work is a preliminary step for using neural networks to speed-up the FSI coupling convergence by providing an accurate initial guess in each time step for classical numerical solvers.

\end{abstract}

\section{Introduction}
In the recent years, machine learning (ML) methods have been widely used in simulation science. The availability of vast amounts of data, either from simulations or experiments, provides the opportunity to adapt data-driven methods to tackle many challenges in this field. For instance, ML methods can be used for reduced order modeling, experimental data processing and shape optimization in fluid mechanics~\cite{brunton2020machine}. As a specific example, ~\cite{vlachas2018data} incorporate long short term memory (LSTM) networks for weather forecasting. Besides, deep neural networks can be used for the parameter calibration of partial differential equations (PDEs). For instance, ~\cite{raissi2019physics} uses the so-called physics informed networks to quantify the unknown parameters in PDEs that are derived for modeling a physical phenomenon.          

In the current work, we aim to incorporate neural networks to solve fluid-structure interaction (FSI) problems. A partitioned approach is followed which divides the problem into two sub-domains according to the occurring physics: 1- fluid domain and 2- structure domain. An individual neural network is trained for each sub-domain to replace the classical solvers. To couple the fluid and the solid solvers, we use the coupling library preCICE~\cite{preCICE}. We employ an implicit coupling scheme at the common boundary, where, for each time step, we iteratively solve both domains until the solution converges. The common interface data from other one will be used as the respective boundary condition for the solver. The data exchange, data mapping and iterative schemes for equation coupling convergence will be provided by the preCICE library. We note, that our ultimate goal is to use the learning framework presented in this work to accelerate the convergence of the FSI coupling. In that case, a dynamic learning strategy can  be applied to train both networks on the fly. The networks use the classical solver's solution for training and produce only a first estimation of the solution for the next time step. This can reduce the number of required iterations. However, this is beyond the scope of current project and is currently under investigation by the authors.      

\begin{figure*}[hbt]
	\centering
	\includegraphics[width=0.9\linewidth]{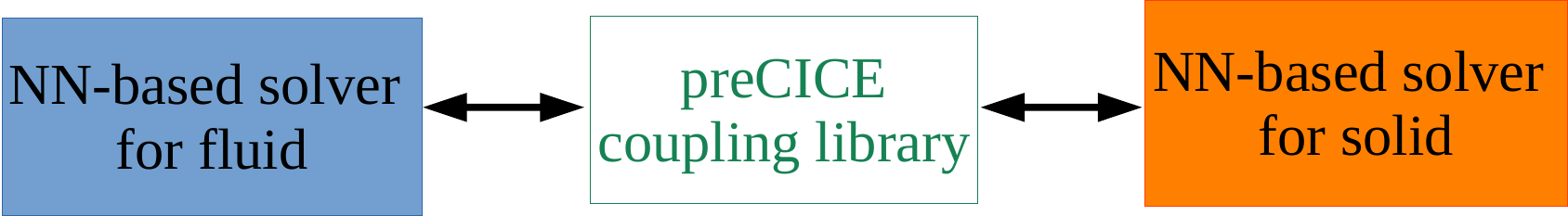}
	\caption{Two neural network based solvers are coupled via the coupling library preCICE.}
	\label{fig:precice}
\end{figure*}

The rest of the paper is organized as follows: Sec.~\ref{sec:math} explains the governing equations for fluid flow, structural deformation and FSI coupling. In Sec.~\ref{sec:nn-arch}, we describe the architecture of the neural network used for each subdomain. In Sec.~\ref{sec:numerics}, we showcase the simulation of the fluid flow inside an elastic tube. We compare the proposed framework's prediction with actual simulation and discuss the results. Finally, Sec.~\ref{sec:conc} summarizes and concludes the paper.

\section{Mathematical models} 
\label{sec:math}
\begin{figure*}[hbt]
	\centering
	\includegraphics[width=0.8\linewidth]{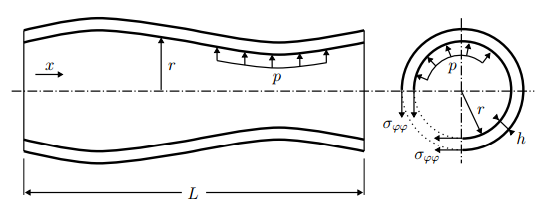}
	\caption{1d tube test case: schematic geometry and parameters~\cite{gatzhammer2014efficient}.}
	\label{fig:tube-1d}
\end{figure*}

We present the governing equations for the simulation of an internal flow inside an elastic tube. The test case is schematically shown in Fig.\ref{fig:tube-1d}. We briefly explain the 1d model equations for the fluid flow, structure deformation and FSI coupling at the common boundary. For more details regarding the test case, please consult~\cite{degroote2008stability}.  

\subsection{Fluid flow}

In this paper, we consider an unsteady and incompressible flow model. Due to the axisymmetric geometry, the flow can be described by quasi-two-dimensional continuity and the momentum equations which read:

\begin{equation} \label{eq:fluid1}
\frac{\partial {\bf a}}{\partial t}+\frac{\partial {\bf(av)}}{\partial x} = 0,
\end{equation}

\begin{equation} \label{eq:fluid2}
\frac{\partial {\bf(av)}}{\partial t}+ \frac{\partial {\bf(av^2)}}{\partial t} + 
\frac{1}{\rho}(\frac{\partial {\bf(ap)}}{\partial x}-p\frac{\partial {\bf a}}{\partial x}) = 0,
\end{equation}

where \textbf{a} is the cross-sectional area (dependent on $x$), \textbf{v} is the flow velocity in $x$-direction, \textbf{p} is the fluid pressure, and $\rho$ is the fluid density. 

\subsection{Structure deformation}
The structure equations for the tube walls are given by a linear elastic constitutive relation law with the scalar circumferential stress
\begin{equation} \label{eq:solid}
\sigma_{\phi \phi} = E\frac{(r-r_0)}{\partial r_0} + \sigma_0,
\end{equation}
where $E$ is the Young's modulus, and $\sigma_0$ the circumferential stress at reference position $r_0$. The motion of the tube wall is, thus, limited to the radial direction.

\subsection{FSI Coupling condition}
The coupling conditions on the fluid-structure interface are derived from the physical equilibrium at the shared boundary. The dynamic FSI interface conditions are given by

\begin{equation} \label{eq:fsi}
pr = \sigma_{\phi \phi} h,
\end{equation}

where $h$ is the thickness of the tube wall. To accelerate the convergence of the FSI coupling, we use a quasi-Newton method, as proposed in~\cite{preCICE}.

\section{Neural network architecture} 
\label{sec:nn-arch}
\begin{figure*}[t]
	\centering
	\includegraphics[width=\textwidth]{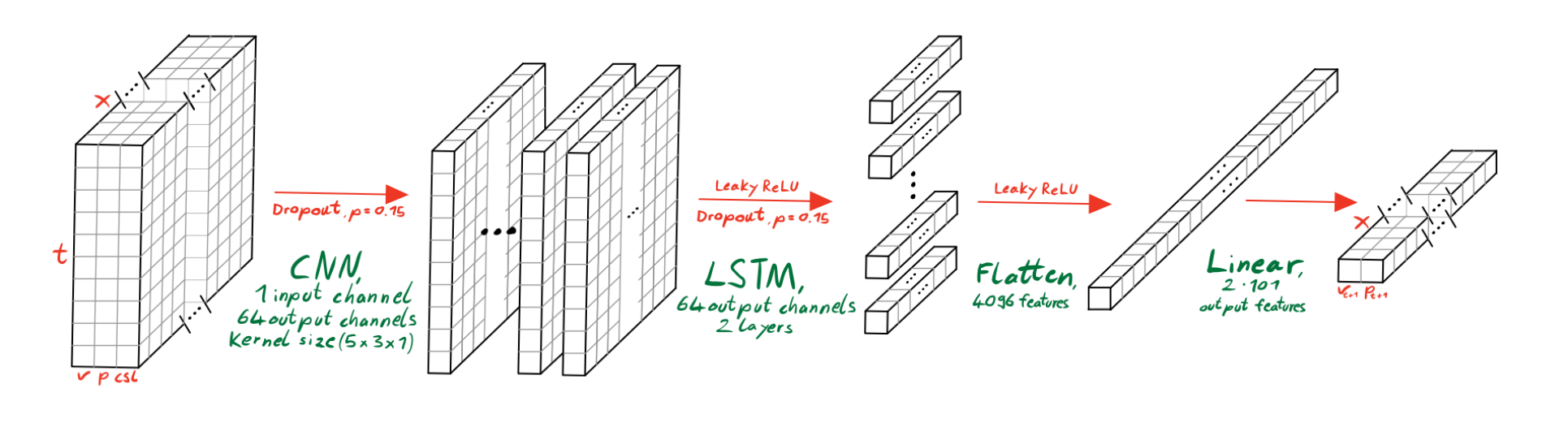}
	\caption{Schematic presentation of the deep neural network. The input consists of domain variables measured at various data points in the domain.}
	\label{fig:layers}
\end{figure*}

In a time-dependent continuum domain, as the test case of the current paper, all unknowns of the system have both spatial and temporal connectivity correlations. Any kind of modeling must account for these connectivity correlations to get accurate results. We use a combination of convolutional (CNN) layers and Long Short Term Memory (LSTM) layers to preserve the mentioned properties. Fig.~\ref{fig:layers} depicts the layers for the fluid solver. We use a similar architecture for the solid solver. The network consists of CNN layers followed by a LSTM layer. At the end, fully connected layers are added to couple various parameters (for example pressure and velocity in fluid solver) to each other. The fluid network uses the pressure, velocity and cross section diameter (from solid solver) as input to predict the pressure and the velocity for the next time steps. On other hand, the solid solver's input consists of the cross section diameter (at the current time step) and the pressure (from the fluid solver) to predict the cross section diameter for the next time steps. 

We use leaky ReLU activation functions~\cite{rect} to map the input of a neuron to its output. A normalized mean square error is considered as the loss function which is given by: 

\begin{equation}
\label{eq:mape}
L := \frac{1}{m}\sum_{k = 1}^{m} \frac{(y_{prediction} - y_{target})^2}{y_{target}^2},
\end{equation}

where $y_{target}$ are the target values and $y_{prediction}$ are the corresponding predictions. The ADAM optimizer~\cite{adam} is incorporated to minimize the loss. We use the dropout technique (as described in~\cite{goodfellow2016deep}) to avoid overfitting during the training process.

\section{Numerical results}
\label{sec:numerics}
In this section, we compare the solution of a numerical simulation with the output of the neural network solver to investigate the accuracy of the proposed framework. The network receives the domain information and the common boundary data at the current time step as input and is expected to predict the future domain status. At the beginning, both networks receive the solution for only 10 time steps to produce the solution at time step 10. Afterwards, the solution at time step 10 is inserted to the input and the networks continue predicting the solution at next time steps. Note, that, the training and validation data are produced using a numerical simulation. 

\begin{figure*}[!htb]
	\centering
	\begin{tabular}{@{}cccc@{}}
		\includegraphics[width=0.23\textwidth]{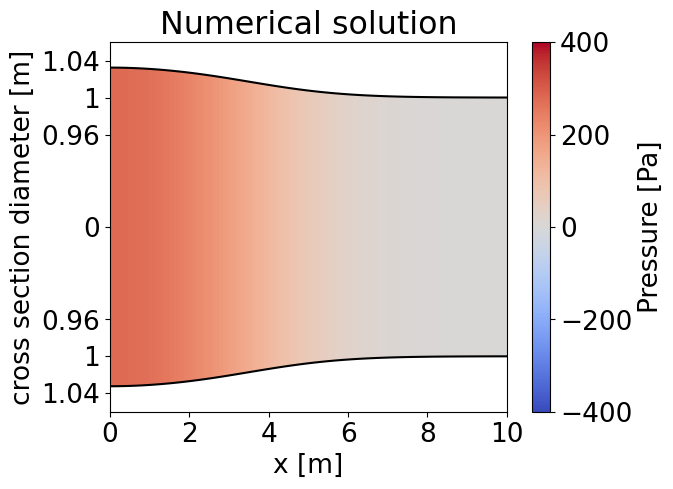} &
		\includegraphics[width=0.23\textwidth]{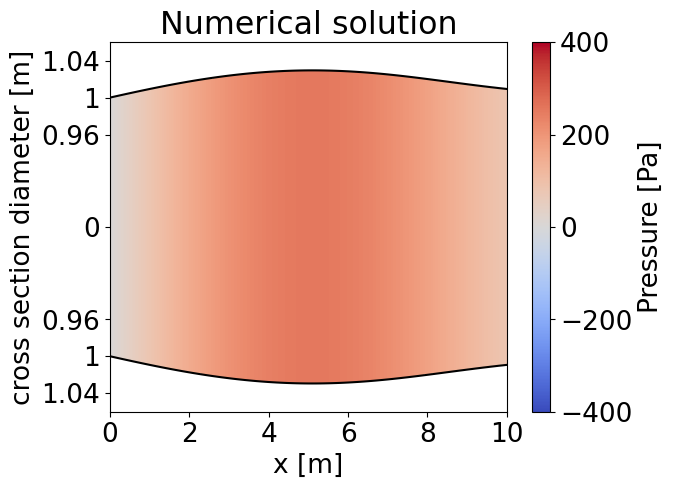} &
		\includegraphics[width=0.23\textwidth]{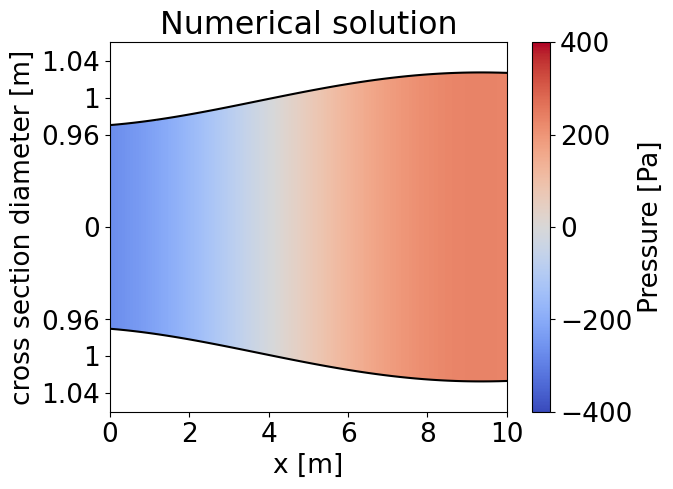} &
		\includegraphics[width=0.23\textwidth]{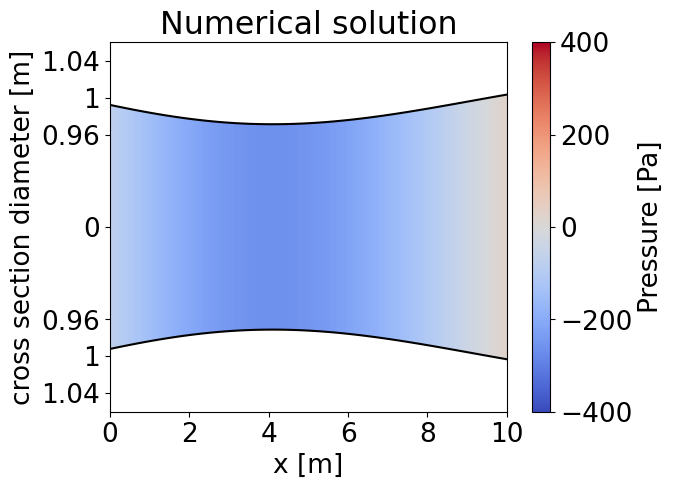}   \\
		\includegraphics[width=0.23\textwidth]{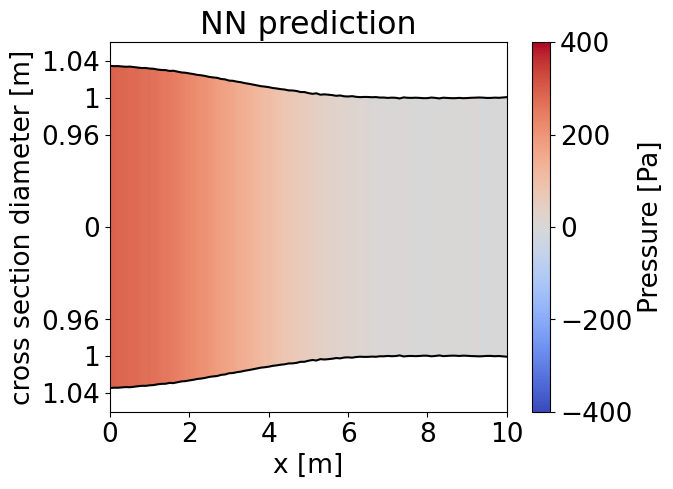} &
		\includegraphics[width=0.23\textwidth]{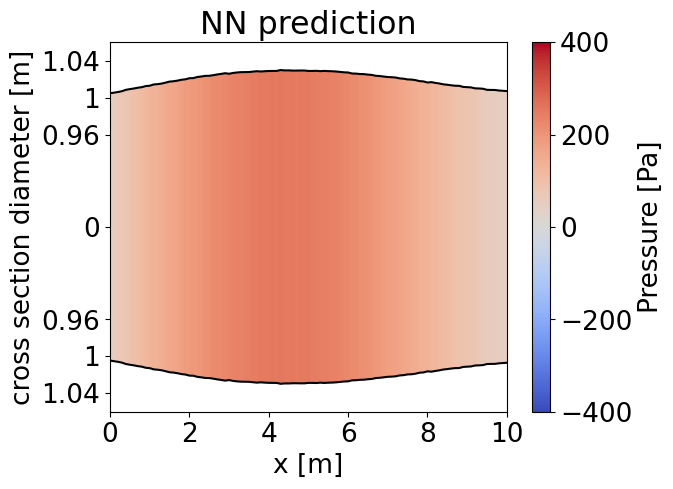} &
		\includegraphics[width=0.23\textwidth]{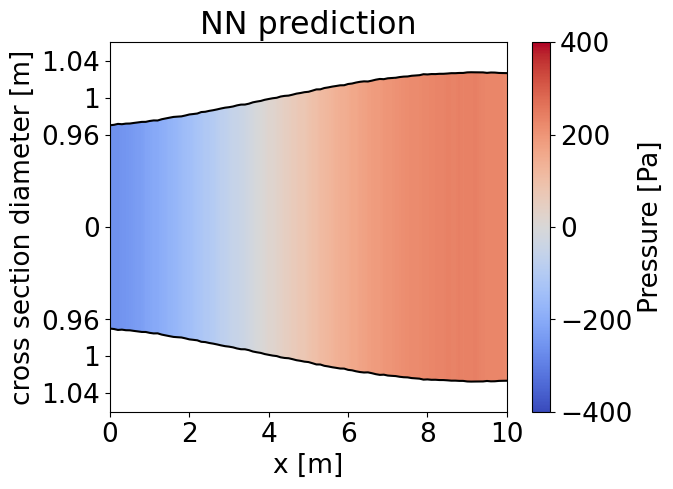} &
		\includegraphics[width=0.23\textwidth]{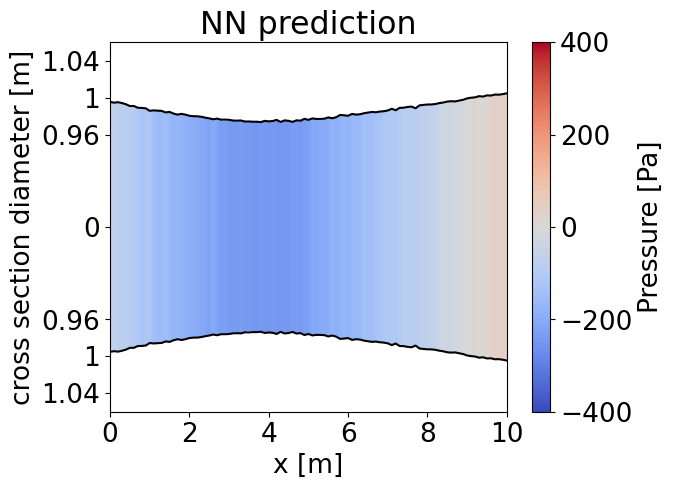}   
	\end{tabular}
	\caption{Comparison of the numerical simulation with the neural network output at $t=$ $0.052$, $0.100$,$0.14$, $0.192$ s (left to right). The upper row corresponds to the numerical solution and the lower row to the neural network prediction.}
	\label{fig:accuracy}
\end{figure*}

Fig.~\ref{fig:accuracy} compares the neural network prediction and the numerical solution for various time steps. The comparison includes both the pressure distribution and the cross section diameter. We observe a very good agreement between the prediction and the simulation data. We see an increasing error in the results, which is due to the accumulative error that comes from using the network prediction as an input for the next predictions. A small error at each prediction can accumulate to a considerable amount after some time steps. However, using an implicit coupling scheme has significantly reduced the error and improved the prediction's accuracy. On the other hand, since the short term predictions are very accurate, almost identical to the simulation results while much cheaper to get, the framework will fulfill the ultimate goal of the project for acceleration of the FSI coupling by producing a fast and accurate first estimation.  

\section{Conclusion}
\label{sec:conc} 
A partitioned scheme for machine learning of coupled fluid-structure interaction problems is introduced, where the proposed method divides the simulation domain into a fluid and a structure domain and incorporates an independent neural network for each. The preCICE coupling library is used for boundary data communication and equation coupling. The neural network for each subdomain consists of CNN, LSTM and fully connected layers to account for both spatial and temporal connectivity correlations. We showcase learning of a 1d fluid flow inside a flexible tube. The input data for each network consists of the domain variables and common boundary values. The inputs of each neuron are mapped to the output using a leaky ReLU activation function. A normalized mean square error function is used along with the ADAM optimization method to calculate the network's weights. 

The numerical accuracy analysis demonstrated a very good agreement between network prediction and simulation data. We are able to predict both the fluid flow and the structural deformation. However, we observed an small accumulative error in the predictions which is due to the tiny errors for each time step prediction which can sum up to a larger amount when long term predictions are desired. However, the short term predictions are very accurate which is sufficient for convergence acceleration in strongly coupled multi-physics simulations.   

\section*{Acknowledgment}
We thank the Deutsche Forschungsgemeinschaft (DFG, German Research Foundation) for supporting this work by funding - EXC2075 – 390740016 under Germany's Excellence Strategy. We acknowledge the support by the Stuttgart Center for Simulation Science (SimTech).

\bibliographystyle{unsrt}
\bibliography{paper}

\begin{thebibliography}{1}

\bibitem{brunton2020machine}
Steven~L Brunton, Bernd~R Noack, and Petros Koumoutsakos.
\newblock Machine learning for fluid mechanics.
\newblock {\em Annual Review of Fluid Mechanics}, 52:477--508, 2020.

\bibitem{vlachas2018data}
Pantelis~R Vlachas, Wonmin Byeon, Zhong~Y Wan, Themistoklis~P Sapsis, and
  Petros Koumoutsakos.
\newblock Data-driven forecasting of high-dimensional chaotic systems with long
  short-term memory networks.
\newblock {\em Proceedings of the Royal Society A: Mathematical, Physical and
  Engineering Sciences}, 474(2213):20170844, 2018.

\bibitem{raissi2019physics}
Maziar Raissi, Paris Perdikaris, and George~E Karniadakis.
\newblock Physics-informed neural networks: A deep learning framework for
  solving forward and inverse problems involving nonlinear partial differential
  equations.
\newblock {\em Journal of Computational Physics}, 378:686--707, 2019.

\bibitem{preCICE}
Hans-Joachim Bungartz, Florian Lindner, Bernhard Gatzhammer, Miriam Mehl,
  Klaudius Scheufele, Alexander Shukaev, and Benjamin Uekermann.
\newblock {preCICE} -- a fully parallel library for multi-physics surface
  coupling.
\newblock {\em Computers and Fluids}, 141:250--258, 2016.
\newblock Advances in Fluid-Structure Interaction.

\bibitem{degroote2008stability}
Joris Degroote, Peter Bruggeman, Robby Haelterman, and Jan Vierendeels.
\newblock Stability of a coupling technique for partitioned solvers in fsi
  applications.
\newblock {\em Computers \& Structures}, 86(23-24):2224--2234, 2008.

\bibitem{gatzhammer2014efficient}
Bernhard Gatzhammer.
\newblock {\em Efficient and flexible partitioned simulation of fluid-structure
  interactions}.
\newblock PhD thesis, Technische Universit{\"a}t M{\"u}nchen, 2014.

\bibitem{rect}
Xavier Glorot, Antoine Bordes, and Yoshua Bengio.
\newblock Deep sparse rectifier neural networks.
\newblock In {\em Proceedings of the fourteenth international conference on
  artificial intelligence and statistics}, pages 315--323, 2011.

\bibitem{adam}
Diederik Kingma and Jimmy Ba.
\newblock Adam: a method for stochastic optimization (2014).
\newblock {\em arXiv preprint arXiv:1412.6980}, 15, 2015.

\bibitem{goodfellow2016deep}
Ian Goodfellow, Yoshua Bengio, Aaron Courville, and Yoshua Bengio.
\newblock {\em Deep learning}, volume~1.
\newblock MIT Press, 2016.

\end{thebibliography}

\end{document}